\documentclass[journal]{IEEEtran}
\usepackage{amsfonts}
\usepackage{amssymb}
\usepackage{amsmath}
\usepackage{algorithm}
\usepackage{multirow}
\usepackage{xcolor}
\usepackage{graphicx}
\usepackage{cite}
\usepackage{exscale}
\usepackage{relsize}
\usepackage{algpseudocode}
\usepackage{graphics}
\usepackage{mathrsfs}
\usepackage{siunitx}
\usepackage{threeparttable}
\usepackage{url}
\usepackage{booktabs}
\usepackage{lipsum}
\usepackage{array}
\DeclareMathOperator*{\argmax}{argmax}
\newcommand{\bm}[1]{\mbox{\boldmath{$#1$}}}

\newtheorem{Def}{Definition}

\ifCLASSOPTIONcompsoc
\usepackage[caption=false,font=normalsize,labelfont=sf,textfont=sf]{subfig}
\else
\usepackage[caption=false,font=footnotesize]{subfig}
\fi

\usepackage[T1]{fontenc}

\title{Optimal Sampling of Water Distribution Network Dynamics using Graph Fourier Transform}

\author{Zhuangkun Wei\textsuperscript{1}, Alessio Pagani\textsuperscript{2}, Guangtao Fu\textsuperscript{2,3}, Ian Guymer\textsuperscript{4}, Wei Chen\textsuperscript{5}, Julie McCann\textsuperscript{2,6}, Weisi Guo\textsuperscript{1,2*}

\thanks{\textsuperscript{1}University of Warwick, UK. \textsuperscript{2}The Alan Turing Institute, UK. \textsuperscript{3}University of Exeter, UK.  \textsuperscript{4}University of Sheffield, UK. \textsuperscript{5}Beijing Jiaotong University, China. \textsuperscript{6}Imperial College, UK.\textsuperscript{*}Corresponding Author: weisi.guo@warwick.ac.uk. }}

\begin{document}

\maketitle

\begin{abstract}
Water Distribution Networks (WDNs) are critical infrastructures that ensure safe drinking water. One of the major threats is the accidental or intentional injection of pollutants. Data collection remains challenging in underground WDNs and in order to quantify its threat to end users, modeling pollutant spread with minimal sensor data is can important open challenge. Existing approaches using numerical optimisation suffer from scalability issues and lack detailed insight and performance guarantees. Applying general data-driven approaches such as compressed sensing (CS) offer limited improvements in sample node reduction. Graph theoretic approaches link topology (e.g. Laplacian spectra) to optimal sensing locations, it neglects the complex dynamics.

In this work, we introduce a novel Graph Fourier Transform (GFT) that exploits the low-rank property to optimally sample junction nodes in WDNs. The proposed GFT allows us to fully recover the full network dynamics using a subset of data sampled at the identified nodes. The proposed GFT technique offers attractive improvements over existing numerical optimisation, compressed sensing, and graph theoretic approaches. Our results show that, on average, with nearly 30-40\% of the junctions monitored, we are able to fully recover the dynamics of the whole network. The framework is useful beyond the application of WDNs and can be applied to a variety of infrastructure sensing for digital twin modeling.
\end{abstract}

\section{Introduction}\label{sec:Intro}

Clean potable water has been described as the blue gold of the 21st century \cite{bluegold} for its importance and scarcity \cite{Mekonnene1500323}. As such, its storage and distribution are fundamental for the welfare of our society. Water distribution is ensured by a complex network of pipes that span over long distances (more than 350,000 km of water pipes in the UK \cite{ukwater-pipes}), connecting reservoirs and tanks to distribution points.
Due to this enormous extent and their underground nature, WDNs are under threats of contamination \cite{Pye713} from a variety of pollution run-off events, both accidental (e.g., pesticide contamination \cite{doi:10.1080/03601239009372674}) or intentional (e.g., terrorist-motivated events \cite{mays04,doi:10.1080/07900620903392158}), potentially affecting hundreds of households.

Water distribution is under increased stress of human demand and drought that arises from climate change. In the UK, it is expected that 4,000 Mega litres/day (26\% increase) of extra water is needed in the near future \cite{NIC18}. Failure to respond to stressors can lead to a $\pounds40bn$ cost in emergency response. It is expected that improving the resilience of water distribution systems will cost $\pounds21bn$, and the primary focus areas include reducing leakage and demand, as well as improving demand management and resilience to stressors (present and future). This is part of wider resilience frameworks (e.g. City Resilience Index - Arup \& Rockefeller Foundation, and Ofwat Towards Resilience) \cite{Welsh17}.

Despite the national importance of WDNs, efforts to fully understand optimal data collection as a function of both the complex network topology and the interconnected internal transport dynamics are still limited and inaccurate, especially when the WDNs face stressors due to incidents or attacks. Installing a sensor in each junction would be the obvious solution to monitor various dynamic states, however this is often not possible because of the high cost \cite{detect-chemical12} and the maintenance difficulty in accessing pipes and junctions buried underground. This raises the necessity of optimized sensor placement \cite{Chang12}, with the objective of reducing the number of sensors in WDNs without hindering the efficiency of contamination detection.

Ideally, an optimal sensor placement would allow to reconstruct and potentially predict the dynamics in the entire WDN monitoring only a subset of junctions (or pipes). Alternatively, to further reduce the number of sensors, an imperfect reconstruction of the dynamics could be accepted if it guarantees high contaminant detection performance (e.g., low time to detect chemical intrusion, low amount of contaminated water consumed or population affected).

\subsection{State-of-the-Art}
\label{sec:stateoftheart}
WDNs are flow-based complex networks with varying topology and heterogeneous dynamic functions. Several studies have been performed trying to optimize sensor placement from different perspectives, and we review them as 3 categories: engineering optimisation, graph-theoretic analysis, and data-driven compression.

\subsubsection{Numerical Optimization Approaches}
In general, rule based multi-objective optimisation considers a number of factors related to both WDN dynamics, as well as accessibility and complexity aspects of the cyber-physical interface \cite{Chang12}. For example, Berry et al. \cite{Berry05} tackled the problem of sensor placement formulation by optimizing the number of sensors that minimize the expected fraction of population at risk from an attack. The approaches include mixed-integer program (MIP), randomized pollution matrix \cite{Kessler98}, and genetic algorithms \cite{Ostfeld04} formulation. However, this problem becomes unfeasible for large-scale networks, especially for various different pollution dynamics. Computational inefficiencies have been tackled for larger WDNs \cite{doi:10.1061/WR.1943-5452.0000001, Krause08Opt}, which for example use a progressive genetic algorithm (PGA) to solve models for large-scale water distribution networks. In one of the most recent works, Another common approach to optimal sensor placement is to construct a multi-objective optimization framework. This gives the capability to reduce the dimensionality of the network through a sensitivity-informed analysis \cite{Fu15} and incorporates uncertainty in the network's demands and Early Winning System operation \cite{SANKARY2017160}. These computational techniques suffer from the lack of explicit relational knowledge between the topological structure and the underlying dynamics with the optimal sampling points.

\subsubsection{Graph-Based Analytical Approaches}
More explicit approaches, that reduce the computation complexity by removing the need of hydraulic simulations \cite{dinardo18, Fu14}, by examining the Graph Spectral Techniques (GSTs) that identify the most influential points on the base of the topological structure of the networks (e.g. via the Laplacian operator). Moreover, similar work also demonstrated that partitioning the WDN in district meter areas offer better monitoring by sensors and protection from contamination \cite{Ciaponi18}. Other approaches to understand critical points include works \cite{pesenson2008sampling,7439829,7208894}.
However, these approaches do not consider the underlying fluid dynamics and assume that the topology dominates. As such, it is important to create an approach that considers both the complex network topology and the pollution signals. Indeed, work on explicit network dynamics that map complex network topology with local dynamics has been progressing from averaged dynamic estimation \cite{Gao16} to node-level precise estimators \cite{Moutsinas18}. More recently, we have mapped optimal sampling of dynamic networks with explicit linearized dynamics with low-dimensionality \cite{Wei19}. However, the challenge with WDNs is that the underlying Navier-Stokes dynamics with variational Reynolds numbers is high dimensional and highly non-linear \cite{Guymer16}. As such, an analysis of the optimal sampling points as a function of both the network topology and the dynamic equations is not possible.

\subsubsection{Data-Driven Compression Approaches}
One approach that considers the data-structure instead of the network topology is the compressed sensing (CS) \cite{Du15CS, McCann15, Xie17CS}. For a matrix data $\mathbf{X}$ of size $N\times K$ with $rank(\mathbf{X})=r$, \cite{5730578} proved that, for all CS methods, the theoretically minimum number of samples needed is $(N+K-r)\times r$, and a nuclear-norm based convex optimization can be used to recover $\mathbf{X}$. In the context of the WDN scenario with $N$ nodes, this means for each time-step $k\in\{1,\cdots,K\}$, an average of $(N+K-r)\times r/K$ sensors are used. However, there are two potential challenges. For one thing, the method in \cite{5730578} did not guarantee an unchanged sensor deployment for different time-steps, therefore may not be quite suitable for WDN surveillance applications. For another, even if other CS schemes\cite{7274320,6566785} can ensure the unchanged sampling nodes for all times-steps, a homogeneous $(N+K-r)\times r/K$ nodes for sampling for all time is still large. We further analyze the performance of CS in Section II and Section IV.

\subsection{Novelty and Contribution}
\label{sec:novelty}

In this work, we suggest a novel sampling method for the networked dynamic signals in WDNs. The idea stems from the graph frequency analysis, whereby a Graph Fourier Transform (GFT) operator (typically the eigenvector matrix of the Laplacian operator \cite{pesenson2008sampling,7439829}) is adopted to compress the data if it belongs to the low-graph frequency space. To sum up, the main contributions of this paper are listed as follows.

(1) As the dynamic signals (e.g. pressure, flow rate, concentration of contaminates) in WDNs consist of highly coupled dynamics, we assume that the aggregate dynamics (i.e. a tensor that represents the $K$ time step dynamics at $N$ junctions) can be represented by the dynamics of a smaller optimal set of junctions ($<N$).

(2) By exploiting this low-rank property in (1), we uncover the graph Fourier basis (operator) that would enable us to determine which set of nodes are optimal to recover the full network's dynamics. Compared with the Laplacian operator (graph structure only) that is extracted from the topology information \cite{dinardo18}, the proposed GFT operator is data-driven, thereby capable of concentrating the networked dynamic signal into the low-frequency region, which makes it possible to characterize the signal via the optimal subset of nodes that belongs to the low-frequency region. Compared with compressed sensing (CS) approaches, we are able to achieve a lower set of nodes at the cost of losing generality. Hence, the novel proposed optimal sensor locations consider both the WDN complex network structure, the underlying data-driven dynamics, and the initial perturbation signal (e.g. chemical pollution at source).

(3) To validate the proposed method, we study the spread of a chemical component in a WDN using the EPANET simulator. The simulation demonstrates that for any $r$-rank dynamic data matrix, a selection of $r$ nodes over the WDN can ensure the full recovery of the chemical propagation over time in all junctions, which has a superior performance compared to \textbf{compressed sensing} (at least $(N+K-r)\times r/K>r$), and the Laplacian based sampling scheme (no guarantees on recovery of dynamics - see results and discussion in Section IV-B). This enables us to inform WDN operators where best to put sensors given a particular perturbation scenario.

\subsection{Organisation}
The rest of paper is structured as follows. In Section II, we describes the nonlinear dynamical WDN system model, and the aim of this paper. In Section III, we elaborate the proposed sampling method. In Section IV, the sampling and recovery performance of the proposed method is evaluated, and the comparison with the traditional Laplacian sampling scheme is provided. In Section V, we conclude the paper and discuss the potential future areas of the research. \\

\section{Model Formulation and Problem Analysis}

In this section, we describe the WDN and the dynamic chemical signal overthe network. Then, two competitive schemes (i.e., the CS scheme, and the Graph sampling sheme based on Laplacian operator).

\subsection{WDN Model}
\begin{figure}[!t]
\centering
\includegraphics[width=3.5in]{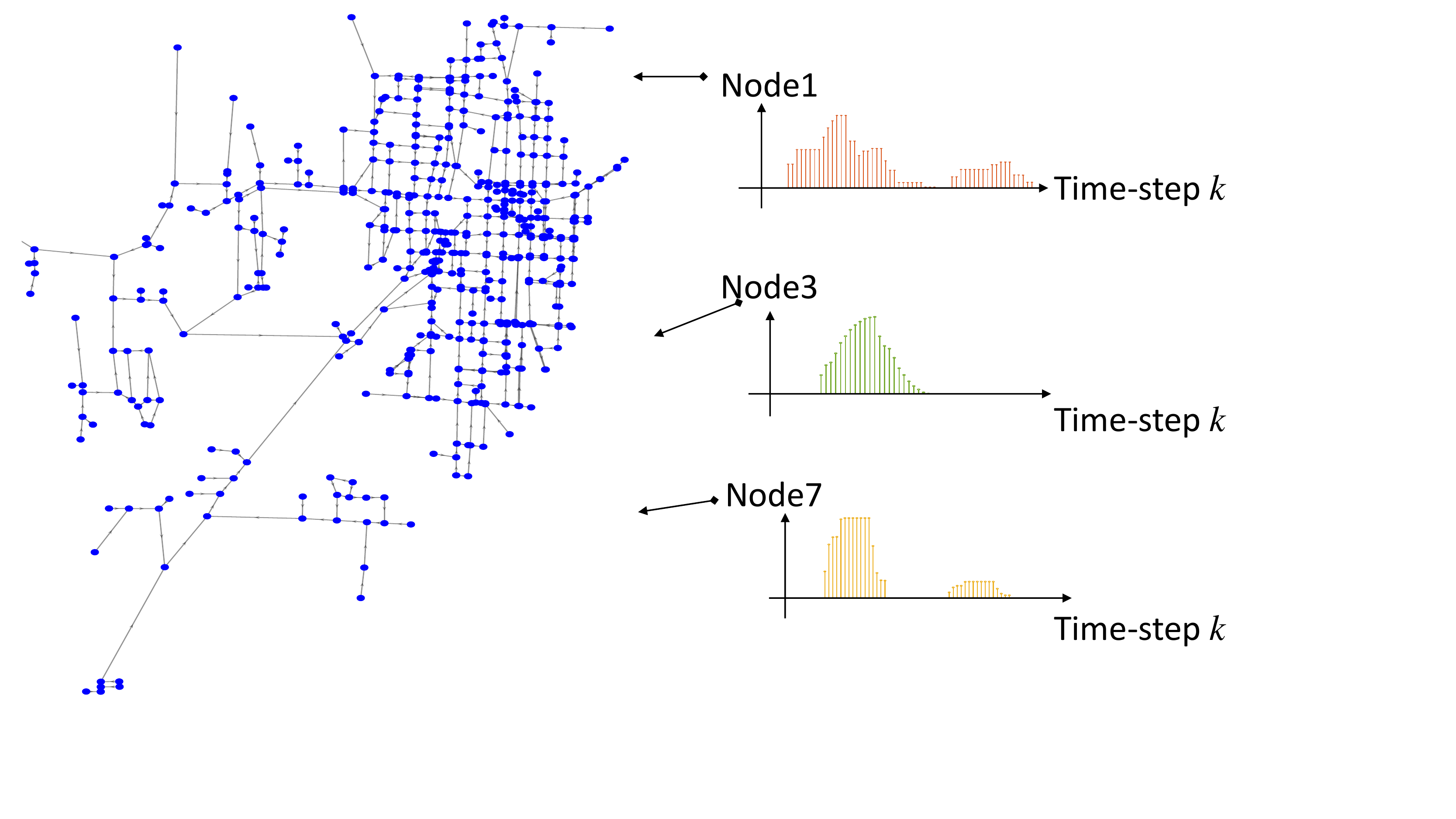}
\caption{Illustration of the WDN and the networked signals. }
\label{fig1}
\end{figure}

The network is configured by a static graph denoted as $G(\mathcal{V}, \mathbf{A})$. $\mathcal{V}=\{1,\cdots,N\}$, $N\in\mathbb{N}^+$ is a set of indices of the total nodes, with different types (e.g., the junction, the reservoir, or the tank \cite{EPANET2}). $\mathbf{A}$ is the adjacency matrix, of which the element $a_{n,m}\in\{0,1\}$ represents an existence ($a_{n,m}=1$) of a directed link from node $m$ to node $n$. The link can be the pump, the valve and the pipe \cite{EPANET2}. For each node $n\in\mathcal{V}$, various types of information (e.g., the water demands, the head-loss, and the water-quality) can be monitored by the sensor if placed on the node. In this paper, we consider the water-quality in terms of the chemical concentration propagated via the network. The WDN topology and the networked chemical data are illustrated in Fig. \ref{fig1}.

The discrete-time chemical data is given as an $N\times K$ matrix $\mathbf{X}=[\mathbf{x}_1,\mathbf{x}_2,\cdots,\mathbf{x}_K]$, where $N=|\mathcal{V}|$ represents the number of nodes in WDN, and $\mathcal{K}=\{1,\cdots,K\}$ is the set of total discrete time-steps. As such, the purpose of this paper is finding a sampling node set, denoted as $\mathcal{S}\subset\mathcal{V}$, such that there exists a recovering matrix denoted as $\bm{\Phi}$:
\begin{equation}
    \mathbf{X}=\bm{\Phi}\cdot\mathbf{X}_{\mathcal{S}\mathcal{K}},
\end{equation}
where $\mathbf{X}_{\mathcal{S}\mathcal{K}}$, the samples of $\mathbf{X}$, has rows with indices in set $\mathcal{S}$, and columns with indices in set $\mathcal{K}$.

\subsection{Two Competitive Schemes}

\subsubsection{Compressed Sensing}
Compressed sensing is a sampling framework to recover sparse signals with a few measurements (or samples). In the context of the WDN signal, the idea is to sparsely represent $\mathbf{X}$ under an $N\times N$ basis $\mathbf{P}$, so that the samples $\mathbf{X}_{\mathcal{S}\mathcal{K}}$ can recover the sparse representation, which subsequently can reconstruct $\mathbf{X}$ \cite{7274320,6566785}. The sampling process is illustrated in Fig \ref{whole}(b). For each time-step $k\in\mathcal{K}$, as we denote $\mathbf{c}_k$ as the sparse representation, $\mathbf{x}_k$ is expressed as:
\begin{equation}
    [\mathbf{x}_1,\mathbf{x}_2,\cdots,\mathbf{x}_K]=\mathbf{P}\cdot[\mathbf{c}_1,\mathbf{c}_2,\cdots,\mathbf{c}_K],
\end{equation}
where $\mathbf{P}$ is an invertible transformation matrix of size $N\times N$, composed by the principal component analysis (PCA) \cite{6287522}. As such, the sampling and recovery issue can be pursued by selecting $\mathcal{S}\subset\mathcal{V}$ such that the restricted isometry property (RIP) is satisfied. That is, the sampling node set $\mathcal{S}\subset\mathcal{V}$ satisfies\footnote{In Eq. (\ref{rip}), in order to use an unchanged $\mathcal{S}$ for all time-steps $1,\cdots,K$, we should consider the maximal sparsity of $[\mathbf{c}_1,\cdots,\mathbf{c}_K]$, i.e., $\gamma=\max_{k\in\mathcal{K}}\|\mathbf{c}_k\|_{l_0}$. Otherwise, if $\gamma<\max_{k\in\mathcal{K}}\|\mathbf{c}_k\|_{l_0}$, the $\mathbf{c}_k,k=\argmax_{k\in\mathcal{K}}\|\mathbf{c}_k\|_{l_0}$ cannot be recovered.}
\begin{equation}
    1-\delta_{2\gamma}\leq\frac{\|\mathbf{P}_{\mathcal{S}\mathcal{V}}\cdot\mathbf{c}\|_{l_2}^2}{\|\mathbf{c}\|_{l_2}^2}\leq 1+\delta_{2\gamma},~\gamma=\max_{k\in\mathcal{K}}\|\mathbf{c}_k\|_{l_0}
    \label{rip}
\end{equation}
for any $2\gamma$ sparse $\mathbf{c}$ and some $\delta_{2\gamma}\in[0,1]$.
Then, as we derive the samples $\mathbf{X}_{\mathcal{S}\{k\}}$, $\mathbf{c}_k$ can be recovered via convex optimization:
\begin{equation}
    \hat{\mathbf{c}}_k=\underset{\mathbf{c}_k\in\mathbb{R}^N}{argmin}\|\mathbf{c}_k\|_{l_1},\text{~such that~} \mathbf{X}_{\mathcal{S}\{k\}}=\mathbf{P}_{\mathcal{S}\mathcal{V}}\cdot\mathbf{c}_k,
\end{equation}
and therefore, $\hat{\mathbf{x}}_k=\mathbf{P}\cdot\hat{\mathbf{c}}_k$, $\hat{\mathbf{X}}=[\hat{\mathbf{x}}_1,\cdots,\hat{\mathbf{x}}_K]$.

\begin{table}
\centering
\caption{Comparison of Size of sampling node set such that RMSE$<10^{-8}$ among different sampling methods.}
\label{table1}
\begin{threeparttable}
\begin{tabular}{lll}
\toprule
\multirow{2}{1cm} {}& \multirow{2}{1cm} {Methods} & {Sampling node set size, s.t. RMSE<$10^{-8}$}   \\
\cline{3-3}
   &    & Data with $rank(\mathbf{X})=r\leq N$\\
  \midrule
  \multirow{2}{1.4cm} {Graph sampling} &  {Data-driven} & {\centering $r$} \\
   & {Laplacian} & $\geq r,~\leq N$ \\
  \multirow{2}{1.4cm} {Compressed sensing} & {DCT basis} & $\geq(N+K-r)r/K\geq r,~\leq N$  \\
   & {PCA basis} & $\geq(N+K-r)r/K\geq r,~\leq N$  \\
  \bottomrule
  \end{tabular}
  \end{threeparttable}
\end{table}

However, it is noteworthy that in order to ensure the RIP in Eq. (\ref{rip}), any $2\gamma$ columns of $\mathbf{P}_{\mathcal{S}\mathcal{V}}$ should be linearly independent, from which \cite{6566785} inferred $|\mathcal{S}|=c\cdot\gamma\log N$, with $c\in[1,4]$. Also, for $\mathbf{X}$ with $rank(\mathbf{X})=r$, \cite{5730578} proves the theoretical minimum number of samples as $(N+K-r)r$. Considering the unchanged selection of $\mathcal{S}$ for all time-steps, we have $|\mathcal{S}|\cdot K\geq (N+K-r)r$, and therefore $|\mathcal{S}|=c\cdot\gamma\log N>(N+K-r)r/K$ nodes are needed for sampling, which is large for selecting sampling nodes in WDNs. We provide the CS needed size of the sampling node set $|\mathcal{S}|$ via Table. 1, and Figs. \ref{comparison}-\ref{set} in Section IV.

\subsubsection{Graph Sampling Theory based on Laplacian}
\begin{figure}[!t]
\centering
\includegraphics[width=3.5in]{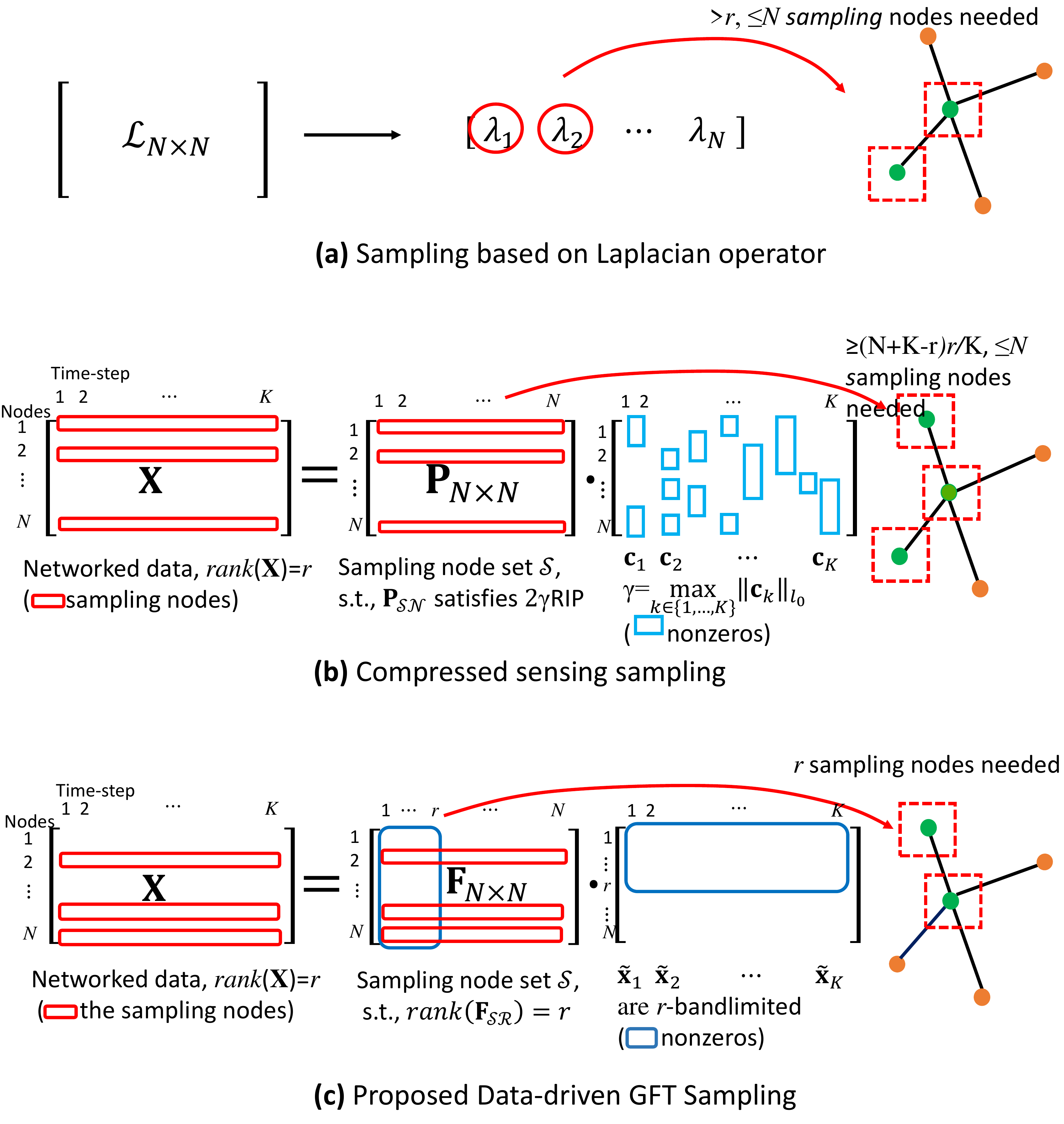}
\caption{Illustration of competitive schemes.}
\label{whole}
\end{figure}

Graph sampling theory samples (compresses) the signal that is bandlimited with respect to a designed graph Fourier transform (GFT) operator, denoted as $\mathbf{F}^{-1}$. Typically, $\mathbf{F}^{-1}$ is constructed via the eigenvectors of the Laplacian operator denoted as $\bm{\mathcal{L}}$, i.e. \cite{pesenson2008sampling,7439829},
\begin{equation}
\begin{aligned}
    \bm{\mathcal{L}}&=\mathbf{D}^{-\frac{1}{2}}\cdot\left(\mathbf{D}-\mathbf{W}\right)\cdot\mathbf{D}^{-\frac{1}{2}}\\
    &=\mathbf{F}\cdot diag\{\lambda_1,\lambda_2,\cdots,\lambda_N\}\cdot\mathbf{F}^{-1},
    \label{eigen-decompose}
\end{aligned}
\end{equation}
where $\mathbf{D}=diag\{d_1,\cdots,d_N\}$ is the degree matrix,  $\lambda_1\leq\lambda_2\leq\cdots\leq\lambda_N$ is the ordered eigenvalues, also referred as the graph frequency (spectral) values ranging from the lowest to the highest parts \cite{pesenson2008sampling,7439829}. In this setting, an $\omega$-bandlimited signal (vector) $\mathbf{x}=[x_1,x_2,\cdots,x_N]^T$ with respect to $\mathbf{F}^{-1}$ is defined to have zero coefficients in the $\mathbf{F}^{-1}$ domain for frequencies above $\omega$, i.e.,
\begin{equation}
    \mathbf{x}=\sum_{i\in\mathcal{N}_{\omega}}\alpha_i\cdot\mathbf{f}_i.
\end{equation}
with $\mathcal{N}_{\omega}=\{i|\lambda_i\leq\omega\}$, and the non-zero coefficient $\alpha_i$. The graph sampling theory states that the $\omega$-bandlimited signal $\mathbf{x}$ can be sampled and fully recovered via a subset of nodes $\mathcal{S}\subset\mathcal{V}$, such that \cite{7439829,7208894}:
\begin{equation}
    rank\left(\mathbf{F}_{\mathcal{S}\mathcal{N}_{\omega}}\right)=|\mathcal{N}_{\omega}|,
    \label{select}
\end{equation}
where $\mathbf{F}_{\mathcal{S}\mathcal{N}_{\omega}}$ denotes the matrix whose rows are indexed via $\mathcal{S}$ and whose columns are indexed via $\mathcal{N}_{\omega}$. The selection in Eq. (\ref{select}) depends on the topology of graph whereby the bandlimited frequencies $\mathcal{N}_{\omega}$ maps to the nodes set $\mathcal{S}$, as is illustrated via Fig. \ref{whole}.(a)

However, directly utilizing the graph sampling theory to identify the sampling node set $\mathcal{S}$ for dynamic WDN signal is challenging. The Laplacian operator cannot ensure that all signals on different time-step (i.e., $\mathbf{X}=[\mathbf{x}_1,\mathbf{x}_2,\cdots,\mathbf{x}_K]$) are $\omega$-bandlimited. This will cause $\mathcal{N}_{\omega}=\{\lambda_1,\lambda_2\cdots,\lambda_N\}$, and inevitably $\mathcal{S}=\mathcal{V}$ (as is shown in Table. 1 and Fig. \ref{comparison}-\ref{set}). In this view, finding an appropriate Fourier operator that enables all $\mathbf{x}_k$ are bandlimited is demanding.

\section{Sampling Process}
In this section, we elaborate our sampling method processed via subset of the nodes to sample and recover the dynamic networked signals on WDNs. In essence, the idea is borrowed from graph sampling theory. We propose a data-driven sampling scheme to (i) generate the GFT operator such that the data $\mathbf{X}$ is bandlimited, (ii) select the optimal sampling set $\mathcal{S}$, and (iii) recover the data via samples from nodes in $\mathcal{S}$.

Before we start, we give the definition on bandlimited matrix signal analogue to the definition of $\omega$-bandlimited vector in graph sampling theory.
\begin{Def}
We say data matrix $\mathbf{X}$ is $r$-bandlimited with respect to an GFT operator $\mathbf{F}^{-1}$, if the rest $N-r$ rows of the frequency response $$\tilde{\mathbf{X}}=\mathbf{F}^{-1}\cdot\mathbf{X}$$ are all zero vectors.
\label{def1}
\end{Def}
\begin{Def}
We call $\mathcal{R}_{\text{cut-off}}=\{1,\cdots,r\}$ the cut-off bandwidth of a data matrix $\mathbf{X}$, if $\mathbf{X}$ is $r$-bandlimited.
\label{def2}
\end{Def}

\subsection{Data-driven GFT Operator}
Given an $N\times K$ data matrix $\mathbf{X}$ with $N$ nodes and $K$ time-steps, the prerequisite of the selection of $\mathcal{S}\subset\mathcal{V}$ enabling full recovery is that $r=rank(\mathbf{X})<N$ \footnote{This is reasonable, because the WDN that consists of $N$ nodes is intrinsically coupled via the fluid dynamics, and therefore, we would expect the rank to be lower than $N$.}. In this view, $\mathbf{X}$ can be transformed into a matrix in which $(N-r)$ rows are $\mathbf{0}$. From Def. \ref{def1}, $\mathbf{X}$ can be viewed as a $r$-bandlimited signal with respect to the transforming matrix. Therefore, this transforming matrix can be used as the GFT operator $\mathbf{F}^{-1}$.

An intuitive way to compute $\mathbf{F}^{-1}$ is to use the maximally linearly independent columns of $\mathbf{X}$, denoted as $\mathbf{x}_{m_1},\cdots,\mathbf{x}_{m_r}$. This is because if the rest $N-r$ rows of $\mathbf{F}^{-1}\cdot[\mathbf{x}_{m_1},\cdots,\mathbf{x}_{m_r}]$ are zero vectors, then every column that can be linearly combined by $\mathbf{x}_{m_1},\cdots,\mathbf{x}_{m_r}$ should be $r$-bandlimited with respect to $\mathbf{F}^{-1}$. By denoting $\mathbf{X}_{\mathcal{V}\mathcal{M}}=[\mathbf{x}_{m_1}.\cdots,\mathbf{x}_{m_r}]$, and the GFT operator $\mathbf{F}^{-1}=[\mathbf{f}_1,\cdots\mathbf{f}_N]^{-1}$, we compute $\mathbf{F}$ via the Schmidt orthogonalization. For $1\leq i\leq r$,
\begin{equation}
\mathbf{f}_i=\frac{\mathbf{x}_{m_i}-\sum_{j=1}^{i-1}\mathbf{f}_j^T\cdot\mathbf{x}_{m_i}\cdot\mathbf{f}_j}{\|\mathbf{x}_{m_i}-\sum_{j=1}^{i-1}\mathbf{f}_j^T\cdot\mathbf{x}_{m_i}\cdot\mathbf{f}_j\|_{l_2}},
\label{feq1}
\end{equation}
with $\mathbf{f}_1=\mathbf{x}_{m_1}/\|\mathbf{x}_{m_1}\|_{l_2}$. Then, for $\mathbf{f}_{r+1},\cdots\mathbf{f}_{N}$, in order to keep them being orthogonal with $\mathbf{f}_1,\cdots,\mathbf{f}_r$, we compute them via the null-space of $[\mathbf{f}_1,\cdots,\mathbf{f}_r]^T$, i.e.,
\begin{equation}
    [\mathbf{f}_1,\cdots,\mathbf{f}_r]^T\cdot\mathbf{y}=\mathbf{0}.
    \label{N-r}
\end{equation}
From Eq. (\ref{N-r}), we derive $N-r$ independent solution vectors $\mathbf{y}_1,\cdots,\mathbf{y}_{N-r}$, each of which is orthogonal with respect to $\mathbf{f}_1,\cdots,\mathbf{f}_r$. Hence, in order to ensure the orthogonality of $\mathbf{F}$, the computation of $\mathbf{f}_{r+1},\cdots,\mathbf{f}_N$ can be pursued via:
\begin{equation}
    \mathbf{f}_{r+i}=\frac{\mathbf{y}_{i}-\sum_{j=1}^{i-1}\mathbf{f}_{r+j}^T\cdot\mathbf{y}_{i}\cdot\mathbf{f}_{r+j}}{\|\mathbf{y}_{i}-\sum_{j=1}^{i-1}\mathbf{f}_{r+j}^T\cdot\mathbf{y}_{i}\cdot\mathbf{f}_{r+j}\|_{l_2}},
    \label{feq2}
\end{equation}
where $1\leq i\leq N-r$. From Eqs. (\ref{feq1})-(\ref{feq2}), the GFT operator $\mathbf{F}^{-1}$ is derived.

With the computation of the GFT operator $\mathbf{F}^{-1}$, we then analyze whether $\mathbf{X}$ is $r$-bandlimited with respect to $\mathbf{F}^{-1}$. We firstly prove that $\mathbf{X}_{\mathcal{V}\mathcal{M}}$ is $r$-bandlimited with respect to $\mathbf{F}^{-1}$ by computing its graph frequency response, denoted as $\tilde{\mathbf{X}}_{\mathcal{V}\mathcal{M}}$, i.e.,
\begin{equation}
    \begin{aligned}
        \tilde{\mathbf{X}}_{\mathcal{V}\mathcal{M}}&=\mathbf{F}^{-1}\cdot\mathbf{X}_{\mathcal{V}\mathcal{M}}\\
        &\overset{\text{(a)}}{=}[\mathbf{f}_1,\cdots,\mathbf{f}_N]^T\cdot[\mathbf{x}_{m_1},\cdots\mathbf{x}_{m_r}]\\
        &\overset{\text{(b)}}{=}\begin{bmatrix}
        \mathbf{R}_{r\times r}\\
        \mathbf{0}_{(N-r)\times r}\\
        \end{bmatrix},
    \end{aligned}
    \label{qr}
\end{equation}
where $\mathbf{R}_{r\times r}$ is an upper-triangular matrix, i.e.,
\begin{equation}
    \mathbf{R}_{r\times r}=\begin{bmatrix}
\mathbf{f}_1^T\cdot\mathbf{x}_{m_1} & \mathbf{f}_1^T\cdot\mathbf{x}_{m_2} & \cdots & \mathbf{f}_1^T\cdot\mathbf{x}_{m_r}\\
 & \mathbf{f}_2^T\cdot\mathbf{x}_{m_2} & \cdots & \mathbf{f}_2^T\cdot\mathbf{x}_{m_r}\\
 & & \ddots & \vdots\\
 & & & \mathbf{f}_r^T\cdot\mathbf{x}_{m_r}
 \end{bmatrix}.
 \label{RR}
\end{equation}
In Eq. (\ref{qr}), (a) holds for fact that the orthogonal $\mathbf{F}$ has $\mathbf{F}^{-1}=\mathbf{F}^T$. (b) is given by $\mathbf{f}_i^T\cdot \mathbf{x}_{m_j}=0$ if $i>j$, since,
\begin{equation}
    \mathbf{f}_i^T\cdot\mathbf{x}_{m_j}=\mathbf{f}_i^T\cdot\sum_{l=1}^{j}c_l\cdot\mathbf{f}_l=\sum_{l=1}^{j}c_l\cdot\left(\mathbf{f}_i^T\cdot\mathbf{f}_l\right)=0,
\end{equation}
where $c_1,\cdots,c_l$ are coefficients.

Then, according to Eq. (\ref{qr}), we can prove that $\mathbf{X}$ is also $r$-bandlimited with respect to $\mathbf{F}^{-1}$, via the computation of its frequency response, denoted as $\tilde{\mathbf{X}}$, i.e.,
\begin{equation}
\begin{aligned}
    \tilde{\mathbf{X}}=&\mathbf{F}^{-1}\cdot\mathbf{X},\\
    \overset{\text{(c)}}{=}&\mathbf{F}^{-1}\cdot\left[\mathbf{X}_{\mathcal{V}\mathcal{M}},~\mathbf{X}_{\mathcal{V}\mathcal{M}}\cdot\bm{\Pi}\right],\\
    \overset{\text{(d)}}{=}&\left[\tilde{\mathbf{X}}_{\mathcal{V}\mathcal{M}},~\tilde{\mathbf{X}}_{\mathcal{V}\mathcal{M}}\cdot\bm{\Pi}\right].
    \label{proof bandlimited}
\end{aligned}
\end{equation}
In Eq. (\ref{proof bandlimited}), (c) holds for that each column of $\mathbf{X}$ can be expressed by the columns from $\mathbf{X}_{\mathcal{V}\mathcal{M}}$ multiplied with an $r\times(K-r)$ matrix $\bm{\Pi}$, since $rank(\mathbf{X}_{\mathcal{V}\mathcal{M}})=rank(\mathbf{X})=r$. (d) indicates that only the first $r$ rows of $\tilde{\mathbf{X}}$ are non-zero, as $\tilde{\mathbf{X}}_{\mathcal{V}\mathcal{M}}$ is the upper triangular matrix with $rank(\tilde{\mathbf{X}}_{\mathcal{V}\mathcal{M}})=r$. From Eq. (\ref{proof bandlimited}), we learn that the derived GFT operator $\mathbf{F}^{-1}$ is the appropriate one that ensures $\mathbf{X}$ is $r$-bandlimited.

\subsection{Selection of Sampling Node Set}
Once we derive the GFT operator $\mathbf{F}^{-1}$ from Eqs. (\ref{feq1})-(\ref{feq2}), we design the selection process of the sampling node set $\mathcal{S}$ that ensures the full recovery. The essence is to find an $\mathcal{S}$ such that reversible transformation between $\mathbf{X}$ and $\mathbf{X}_{\mathcal{S}\mathcal{K}}$ exists.

To do so, we consider the frequency response as the intermediate, i.e., we try to find the reversible computations between $\mathbf{X}$ and $\tilde{\mathbf{X}}_{\mathcal{R}\mathcal{K}}$, and $\tilde{\mathbf{X}}_{\mathcal{R}\mathcal{K}}$ and $\mathbf{X}_{\mathcal{S}\mathcal{K}}$ respectively. Here, $\mathcal{R}$ is a sampling bandwidth that selects the $|\mathcal{R}|$ non-zero rows of $\tilde{\mathbf{X}}$. The illustration of node selection is shown in Fig. \ref{whole}(c).

We firstly analyze the computations between $\mathbf{X}$ and $\tilde{\mathbf{X}}_{\mathcal{R}\mathcal{K}}$. Given that $\mathbf{X}$ is $r$-bandlimited with respect to $\mathbf{F}^{-1}$, the cut-off bandwidth of $\mathbf{X}$ is $\mathcal{R}_{\text{cut-off}}=\{1,\cdots,r\}$, as only the first $r$ rows of $\tilde{\mathbf{X}}$ are non-zero. Therefore, the sampling bandwidth $\mathcal{R}$ can be assigned as:
\begin{equation}
    \mathcal{R}=\mathcal{R}_{\text{cut-off}}.
    \label{15}
\end{equation}
As such we can extract the non-zero frequency response, and in turn compute the original data as:
\begin{equation}
    \tilde{\mathbf{X}}_{\mathcal{R}\mathcal{K}}=\mathbf{F}_{\mathcal{V}\mathcal{R}}^T\cdot\mathbf{X},
    \label{eq17}
\end{equation}
\begin{equation}
    \mathbf{X}=\mathbf{F}_{\mathcal{V}\mathcal{R}}\cdot\tilde{\mathbf{X}}_{\mathcal{R}\mathcal{K}},
    \label{bandlimited}
\end{equation}
in which the reversible computation between $\mathbf{X}$ and $\tilde{\mathbf{X}}_{\mathcal{R}\mathcal{K}}$ is found.

Then, we consider the connection between $\tilde{\mathbf{X}}_{\mathcal{R}\mathcal{K}}$ and $\mathbf{X}_{\mathcal{S}\mathcal{K}}$. For any selection $\mathcal{S}\subset\mathcal{V}$, an $\mathbf{X}_{\mathcal{S}\mathcal{K}}$ can be derived via Eq. (\ref{bandlimited}):
\begin{equation}
    \mathbf{X}_{\mathcal{S}\mathcal{K}}=\mathbf{F}_{\mathcal{S}\mathcal{R}}\cdot\tilde{\mathbf{X}}_{\mathcal{R}\mathcal{K}}.
    \label{14}
\end{equation}
We can infer from Eq. (\ref{14}) that $rank(\mathbf{X}_{\mathcal{S}\mathcal{K}})\leq\text{min}\{rank(\mathbf{F}_{\mathcal{S}\mathcal{R}}),rank(\tilde{\mathbf{X}}_{\mathcal{R}\mathcal{K}})\}$. In order to ensure a reversible computation, we need $\mathbf{F}_{\mathcal{S}\mathcal{R}}$ to be full column rank, i.e.,
\begin{equation}
    rank\left(\mathbf{F}_{\mathcal{S}\mathcal{R}}\right)=|\mathcal{R}|.
    \label{selectnew}
\end{equation}
Therefore, the inverse computation from $\mathbf{X}_{\mathcal{S}\mathcal{K}}$ to $\tilde{\mathbf{X}}_{\mathcal{R}\mathcal{K}}$ can be pursued by multiplying $\mathbf{F}_{\mathcal{S}\mathcal{R}}^T$ from both sides of Eq. (\ref{14}), i.e.,
\begin{equation}
    \tilde{\mathbf{X}}_{\mathcal{R}\mathcal{K}}=(\mathbf{F}_{\mathcal{S}\mathcal{R}}^T\cdot\mathbf{F}_{\mathcal{S}\mathcal{R}})^{-1}\cdot\mathbf{F}_{\mathcal{S}\mathcal{R}}^T\cdot\mathbf{X}_{\mathcal{S}\mathcal{K}}.
    \label{20}
\end{equation}

As such, given by Eqs. (\ref{15})-(\ref{20}), we build the reversible computation between the signal $\mathbf{X}$ and the samples $\mathbf{X}_{\mathcal{S}\mathcal{K}}$, under conditions of Eq. (\ref{15}) and Eq. (\ref{selectnew}). The intuitive description of Eq. (\ref{15}) and Eq. (\ref{selectnew}) is given as follows. For any $r$-bandlimited signal $\mathbf{X}$ with respect to $\mathbf{F}^{-1}$, the sampling bandwidth $\mathcal{R}$ should at least embrace the cut-off $\mathcal{R}_{\text{cut-off}}$, so that the information from the $\mathbf{F}^{-1}$ domain will not lose. In other words, the reversible computation between the signal $\mathbf{X}$ and the frequency response $\tilde{\mathbf{X}}_{\mathcal{R}\mathcal{K}}$ exists. Then, Eq. (\ref{selectnew}) builds the reversible transform between the frequency response $\tilde{\mathbf{X}}_{\mathcal{R}\mathcal{K}}$ and the sampled data $\mathbf{X}_{\mathcal{S}\mathcal{K}}$, which combined with Eq. (\ref{15}) ensures the full recovery.

\subsection{Signal Recovery}
With the help of the sampling node set $\mathcal{S}$, we can sample the data $\mathbf{X}$ via $\mathcal{S}$ and derive the sampling data as $\mathbf{X}_{\mathcal{S}\mathcal{K}}$. By combining Eq. (\ref{bandlimited}) and Eq. (\ref{20}), we compute the recovered data, denoted as $\hat{\mathbf{X}}$ as follows:
\begin{equation}
    \hat{\mathbf{X}}=\mathbf{F}_{\mathcal{V}\mathcal{R}}\cdot (\mathbf{F}_{\mathcal{S}\mathcal{R}}^T\cdot\mathbf{F}_{\mathcal{S}\mathcal{R}})^{-1}\cdot\mathbf{F}_{\mathcal{S}\mathcal{R}}^T\cdot\mathbf{X}_{\mathcal{S}\mathcal{K}}.
    \label{recover}
\end{equation}

\subsection{Sampling Algorithm Flow}
After explaining the design of the sampling method, we provide two algorithm flows for sampling and recovering respectively.

The sampling method is illustrated in Algo. 1. The input is the networked data $\mathbf{X}$ that is waiting to be sampled. Step 1 is to find the maximally linearly independent column vectors $\mathbf{X}_{\mathcal{V}\mathcal{M}}$ from $\mathbf{X}$. Step 2 is to compute the part of the inverse GFT operator, as $\mathbf{F}_{\mathcal{V}\mathcal{R}}=[\mathbf{f}_1,\cdots,\mathbf{f}_r]$. Step 3-7 aims to select the sampling node set $\mathcal{S}$ that is subjected to Eq. (\ref{selectnew}). From Eq. (\ref{selectnew}), we can notice that there are various selections of $\mathcal{S}$. In order to achieve a robust sampling scheme on nodes, we consider the selection of $\mathcal{S}$ that maximizes the minimum singular of $\mathbf{F}_{\mathcal{S}\mathcal{R}}$. As we denote the smallest singular value as $\sigma_{\text{min}}$, we can write the optimal selection in Eq. (\ref{s_option}), i.e.,
\begin{equation}
    \mathcal{S}_{\text{opt}}=\argmax_{\mathcal{S}\subset\mathcal{V}}\sigma_{\text{min}}\left(\mathbf{F}_{\mathcal{S}\mathcal{R}}\right).
    \label{s_option}
\end{equation}
Then, a greedy algorithm is used to realize Eq. (\ref{s_option}) in the form of Step 4-7. Step 8 is to derive the sampled data $\mathbf{X}_{\mathcal{S}\mathcal{K}}$.

\begin{algorithm}[t]
\caption{Sampling Method}
\begin{algorithmic}[1]
\Require
Networked data $\mathbf{X}$
\State Find maximally independent vectors $\mathbf{X}_{\mathcal{V}\mathcal{M}}$ from $\mathbf{X}$.
\State Compute $\mathbf{F}_{\mathcal{V}\mathcal{R}}$ via Eq. (\ref{feq1}).
\State Initialize $|\mathcal{S}|=0$, and $r=rank(\mathbf{X})$.
\While{$|\mathcal{S}|<r$}
\State $n=\argmax_{i}\sigma_{\text{min}}\left(\mathbf{F}_{\left(\mathcal{S}+\{i\}\right)\mathcal{R}}\right)$
\State $\mathcal{S}=\mathcal{S}\cup\{n\}$
\EndWhile
\State Sample $\mathbf{X}$, and derive sampled data $\mathbf{X}_{\mathcal{S}\mathcal{K}}$.
\Ensure
Sampled data $\mathbf{X}_{\mathcal{S}\mathcal{K}}$, part of the inverse GFT operator $\mathbf{F}_{\mathcal{V}\mathcal{R}}$, and the sampling node set $\mathcal{S}$.
\end{algorithmic}
\end{algorithm}

\begin{algorithm}[t]
\caption{Recovery Process.}
\begin{algorithmic}[1]
\Require
Sampled data $\mathbf{X}_{\mathcal{S}\mathcal{K}}$, part of inverse GFT operator $\mathbf{F}_{\mathcal{V}\mathcal{R}}$, and sampling node set $\mathcal{S}$.
\State Compute $\mathbf{F}_{\mathcal{S}\mathcal{R}}$ by selecting the rows of $\mathbf{F}_{\mathcal{V}\mathcal{R}}$ whose indexes belong to $\mathcal{S}$.
\State Compute the recovered data $\tilde{\mathbf{X}}$ via Eq. (\ref{recover}).
\Ensure
The recovered data $\tilde{\mathbf{X}}$.
\end{algorithmic}
\end{algorithm}

The recovery process is provided by Algo. 2. The input is the sampled data $\mathbf{X}_{\mathcal{S}\mathcal{K}}$, part of the inverse GFT operator $\mathbf{F}_{\mathcal{V}\mathcal{R}}$, and the sampling node set $\mathcal{S}$. Step 1 is to compute the (generalized) inverse matrix of $\mathbf{F}_{\mathcal{S}\mathcal{R}}$. Step 2 is to compute the recovered data $\hat{\mathbf{X}}$. \\

\section{Results}
In the following analysis, the performance of our proposed sampling method will be evaluated. First, we analyze the recovery performance via two aspects, i.e., the sampling bandwidth $|\mathcal{R}|$, and the size of the sampling node set $|\mathcal{S}|$. Second, we compare the recovery performances between our proposed sampling method, and the sampling scheme based on Laplacian operator. The recovery performance is measured in terms of the root mean square error (RMSE) of the recovered data $\hat{\mathbf{X}}$, i.e.,
\begin{equation}
\text{RMSE}=\mathbb{E}\{\hat{\mathbf{X}}-\mathbf{X}\}=\sqrt{\frac{1}{NK}\sum_{k=1}^{K}\|\hat{\mathbf{x}}_k-\mathbf{x}_k\|_{l_2}^2}.
\label{rmse_eq}
\end{equation}

The simulations in this work are conducted using the Python package Water Network Tool for Resilience (WNTR) based on EPANET2 \cite{EPANET2}, which is capable of
performing extended-period simulation of hydraulic and water-quality behaviour within pressurizes pipe networks. The simulations are executed on Microsoft Azure \cite{azure}.
The WDN network is configured as $N=102$ nodes, including 100 junctions and 2 reservoirs (as illustrated in Fig. \ref{oneselectnode}(a)). For each junction, a random and unknown water-demand is used. The links are pipes with unknown pressures. We simulate 100 different time-varying chemical contaminant propagated via the WDN. Each data $\mathbf{X}$ with a different perturbation are simulated for 3 hours in $K=168$ time-steps.

\subsection{Influences on Recovery Performance}

\begin{figure*}[!t]
\centering
\includegraphics[width=7in]{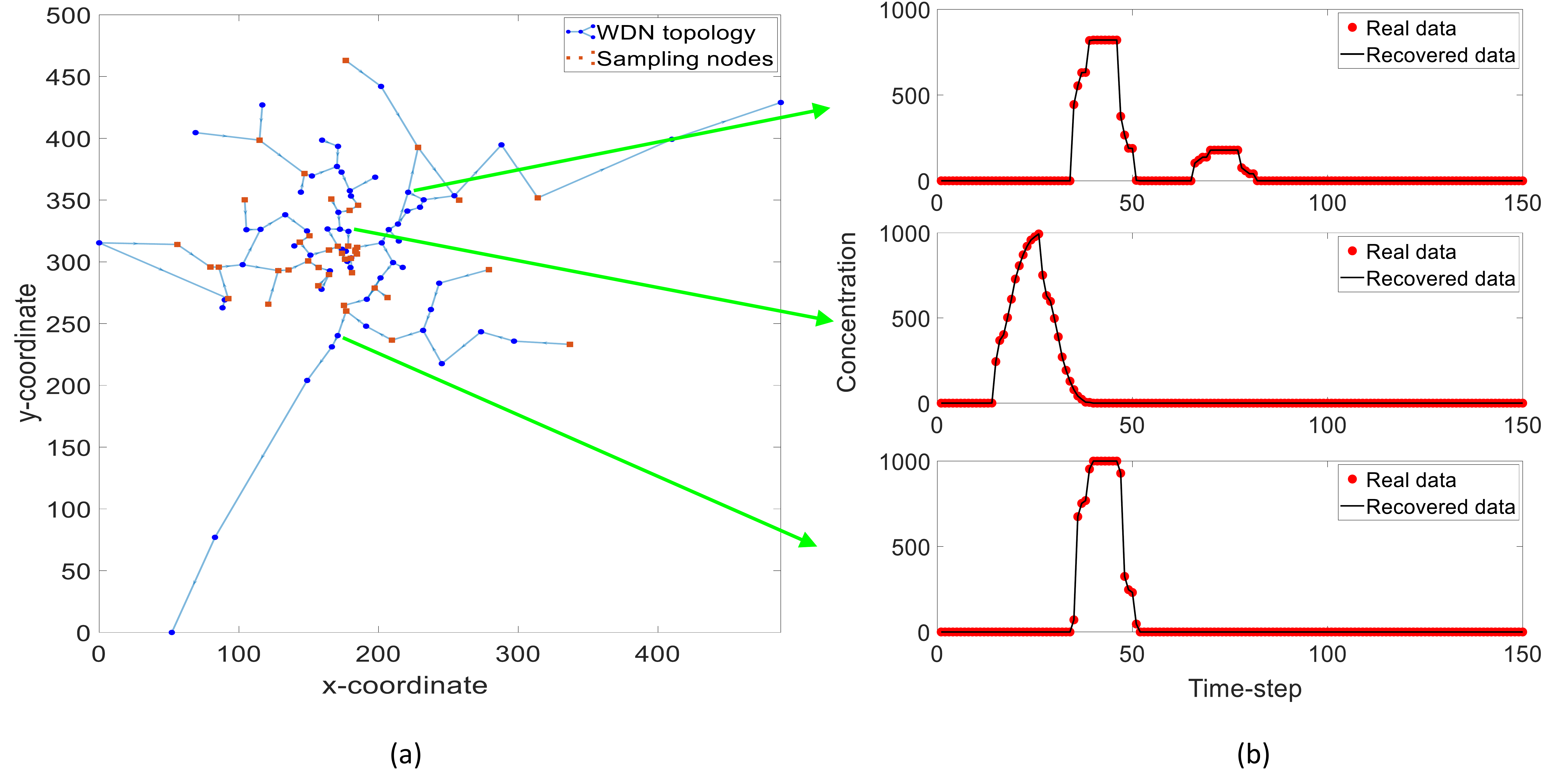}
\caption{Illustration of a networked dynamic data in a WDN, with its sampling and recovery performance. (a) shows the topology of the WDN and the selected sampling nodes. (b) presents 3 examples of real and recovered data from 3 un-sampled nodes. }
\label{oneselectnode}
\end{figure*}

We firstly analyze the recovery performance of our sampling method with respect to the sampling bandwidth $|\mathcal{R}|$, and the size of the sampling node set $|\mathcal{S}|$. One illustration of the sampling and recovery is provided in Fig. \ref{oneselectnode}, whereby Fig. \ref{oneselectnode}(a) show the topology and the selected sampling nodes, and Fig. \ref{oneselectnode}(b) presents the comparisons between real data and the recovered data on 3 un-sampled nodes. In this illustration, we assign $|\mathcal{R}|=|\mathcal{S}|=r$, as suggested in the sampling method (i.e., Eq. (\ref{15}), and Eq. (\ref{selectnew}). We figure out that the perfect recovery is achieved.

\begin{figure*}[!t]
\centering
\includegraphics[width=7in]{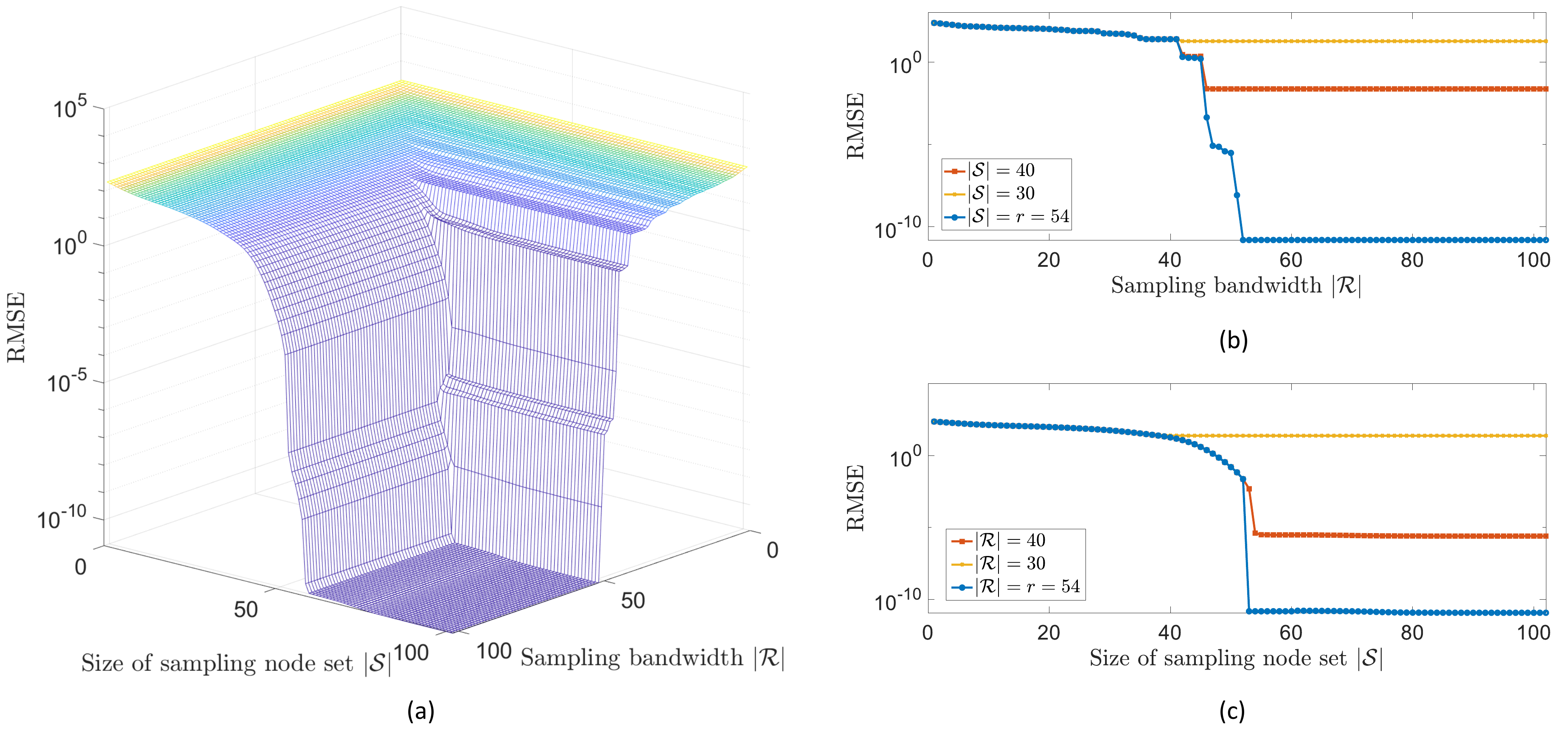}
\caption{The RMSE of the recovered chemical signal, with respect to the sampling bandwidth $|\mathcal{R}|$ and the size of the sampling node set $|\mathcal{S}|$. Sub-plots: (a) is the overall relationship. (b) and (c) show the two planes as we fix $|\mathcal{S}|$ and $|\mathcal{R}|$ respectively.}
\label{rmse}
\end{figure*}

Then, we consider the changes of both the sampling bandwidth $|\mathcal{R}|$ and the size of the sampling node set $|\mathcal{S}|$. Seen from Fig. \ref{rmse}(a), at first, the RMSE decreases with both the increases of $|\mathcal{R}|$ and $|\mathcal{S}|$. Then, after $|\mathcal{R}|$ and $|\mathcal{S}|$ reach the conditions provided from Eq. (\ref{15}), and Eq. (\ref{selectnew} (i.e., $|\mathcal{R}|=|\mathcal{S}|=54$), the RMSE becomes unchanged. The reasons will be discussed as we analyze the Fig. \ref{rmse}(b)-(c).

\subsubsection{Bandwidth of Sampling}
Fig. \ref{rmse}(b) plots the recovery performance influenced by the sampling bandwidth $|\mathcal{R}|$, with 3 fixed sizes of sampling node set (e.g., $|\mathcal{S}|=30,40,54$). It is firstly seen that the RMSEs have obvious differences as different $|\mathcal{S}|$ are considered. For instance, in the case $|\mathcal{S}|=54$, the RMSE keeps lower as opposed to the values from $|\mathcal{S}|=30,40$. This is because with the increase of $|\mathcal{S}|$, more nodes will be sampled for data recovery, which leads to a better recovery performance.

Secondly, we can observe that for each $|\mathcal{S}|$, the RMSE becomes lower as $|\mathcal{R}|$ grows to the rank (i.e., $r=54$), and then remains unchanged when $|\mathcal{R}|>r=54$. We explain the reasons for the two cases respectively. In the case of $|\mathcal{R}|<|\mathcal{R}_{\text{cut-off}}|=r$, the signal from $\mathbf{F}^{-1}$ domain is under-sampled, which further gives rise to the failure of the full recovery. This can be also explained as the loss of the reversible computation between the original data $\mathbf{X}$ and the frequency response selected by the sampling bandwidth $\mathcal{R}$, i.e.,  $\tilde{\mathbf{X}}_{\mathcal{R}\mathbf{K}}$ when $|\mathcal{R}|<|\mathcal{R}_{\text{cut-off}}|$, as Eq. (\ref{bandlimited}) holds no more. In this situation, even if the computation between $\tilde{\mathbf{X}}_{\mathcal{R}\mathbf{K}}$ and the sampled data $\mathbf{X}_{\mathcal{S}\mathcal{K}}$ may exist (e.g., $rank(\mathbf{F}_{\mathcal{S}\mathcal{R}})=|\mathcal{R}|<r$), we still cannot fully recover $\mathbf{X}$ from $\mathbf{X}_{\mathcal{S}\mathcal{K}}$. By contrast, for the case $|\mathcal{R}|\geq r=54$, the total information from $\mathbf{F}^{-1}$ domain remains, and the reversible computation between $\mathbf{X}$ and $\tilde{\mathbf{X}}_{\mathcal{R}\mathbf{K}}$ can be ensured, so the recovery performance depends only on the selection of the sampling nodes (i.e., the fixed $\mathcal{S}$ makes RMSE unchanged).

\subsubsection{Size of Sampling Node Set}
Fig. \ref{rmse}(c) illustrates the recovery performance affected by the size of the sampling node set $|\mathcal{S}|$, with 3 fixed sampling bandwidth (e.g., $|\mathcal{R}|=30, 40, 54$). We can firstly see that the RMSE with a larger fixed $|\mathcal{R}|$ keeps smaller (e.g., the RMSE with $|\mathcal{R}=54|$ is lower than the one with $|\mathcal{R}|=40$). This is due to the reason mentioned above that the larger $|\mathcal{R}|$ can embrace more frequency information from the $\mathbf{F}^{-1}$ domain, which subsequently leads to a better recovery.

Furthermore, we notice that for each fixed sampling bandwidth $|\mathcal{R}|$, the RMSE decreases at first as $|\mathcal{S}|$ grows from $0$ to $r=54$. Then, it remains stable after $|\mathcal{S}|>r=54$. This is because more sampling nodes will improve the recovery performance, and the full recovery can be achieved with the $|\mathcal{R}|=r$, and $|S|\geq r$. Intriguingly, we should also notice that with an under-sampled bandwidth (i.e., $|\mathcal{R}|<|\mathcal{R}_{\text{cut-off}}|$), even if the number of sampling nodes is increasing, the performance will not change after $|\mathcal{S}|>r$. This is because the proposed sampling method is based on the $\mathbf{F}^{-1}$ domain intermediate i.e., the frequency response $\tilde{\mathbf{X}}_{\mathcal{R}\mathbf{K}}$; the loss of information blocks the inverse computation of the data matrix $\mathbf{X}$ from $\tilde{\mathbf{X}}_{\mathcal{R}\mathbf{K}}$, thereby hindering the full recovery from the sample $\mathbf{X}_{\mathcal{S}\mathcal{K}}$ to $\mathbf{X}$.

\subsection{Performance Comparisons}

\begin{figure*}[!t]
\centering
\subfloat[]{\includegraphics[width=3.5in]{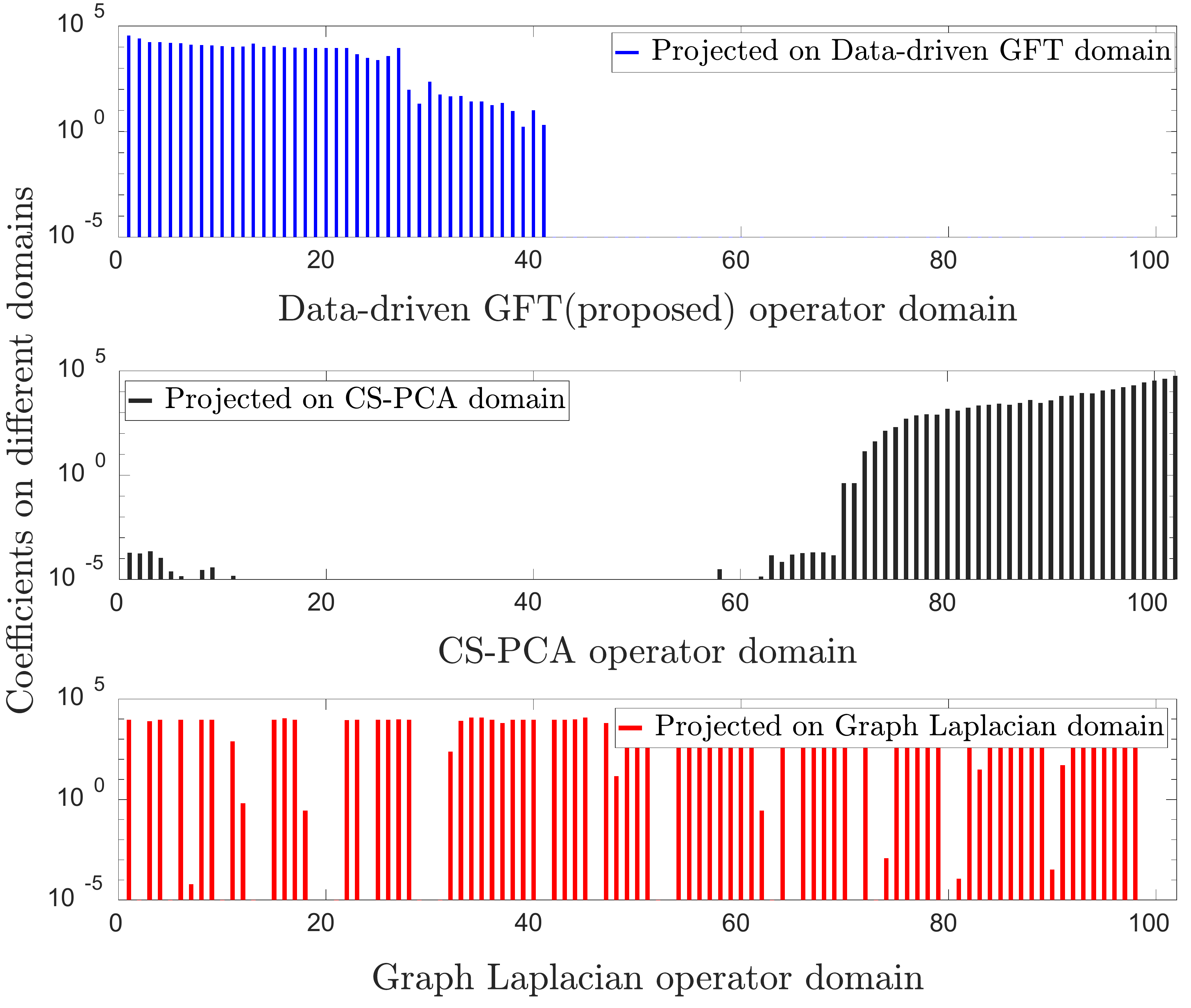}}
\hfil
\subfloat[]{\includegraphics[width=3.5in]{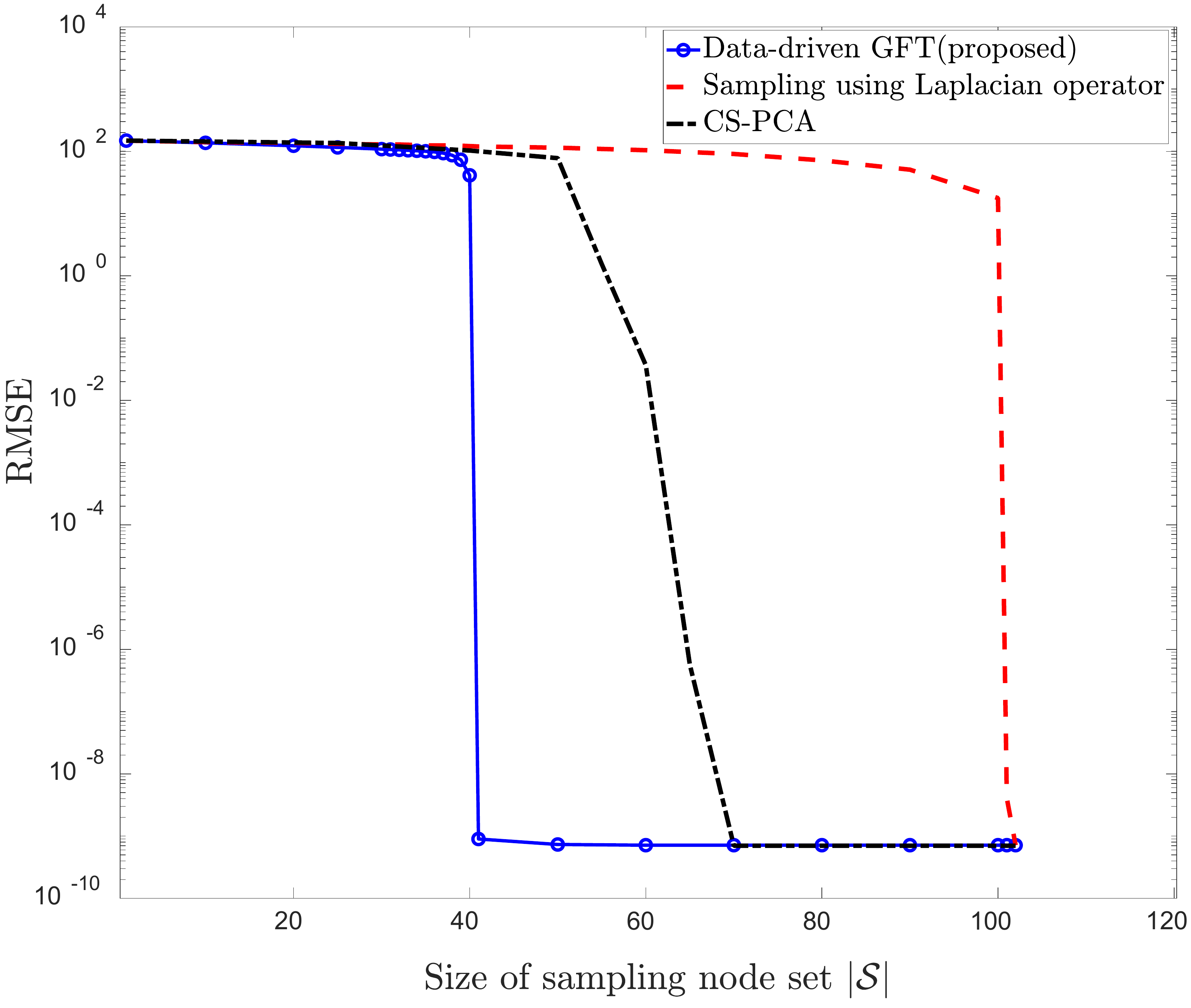}}
\caption{Comparison between the proposed Data-driven GFT sampling method, the sampling based on Laplacian operator, and the compressed sensing with PCA method. (a) gives the frequency response with respect to the proposed GFT operator, the Laplacian operator, and the PCA operator (basis) from the compressed sensing respectively. (b) shows the recovery performance in terms of RMSE varied from the size of sampling node set (i.e., $|\mathcal{S}|$).}
\label{comparison}
\end{figure*}

\begin{figure*}[!t]
\centering
\subfloat[]{\includegraphics[width=3.5in]{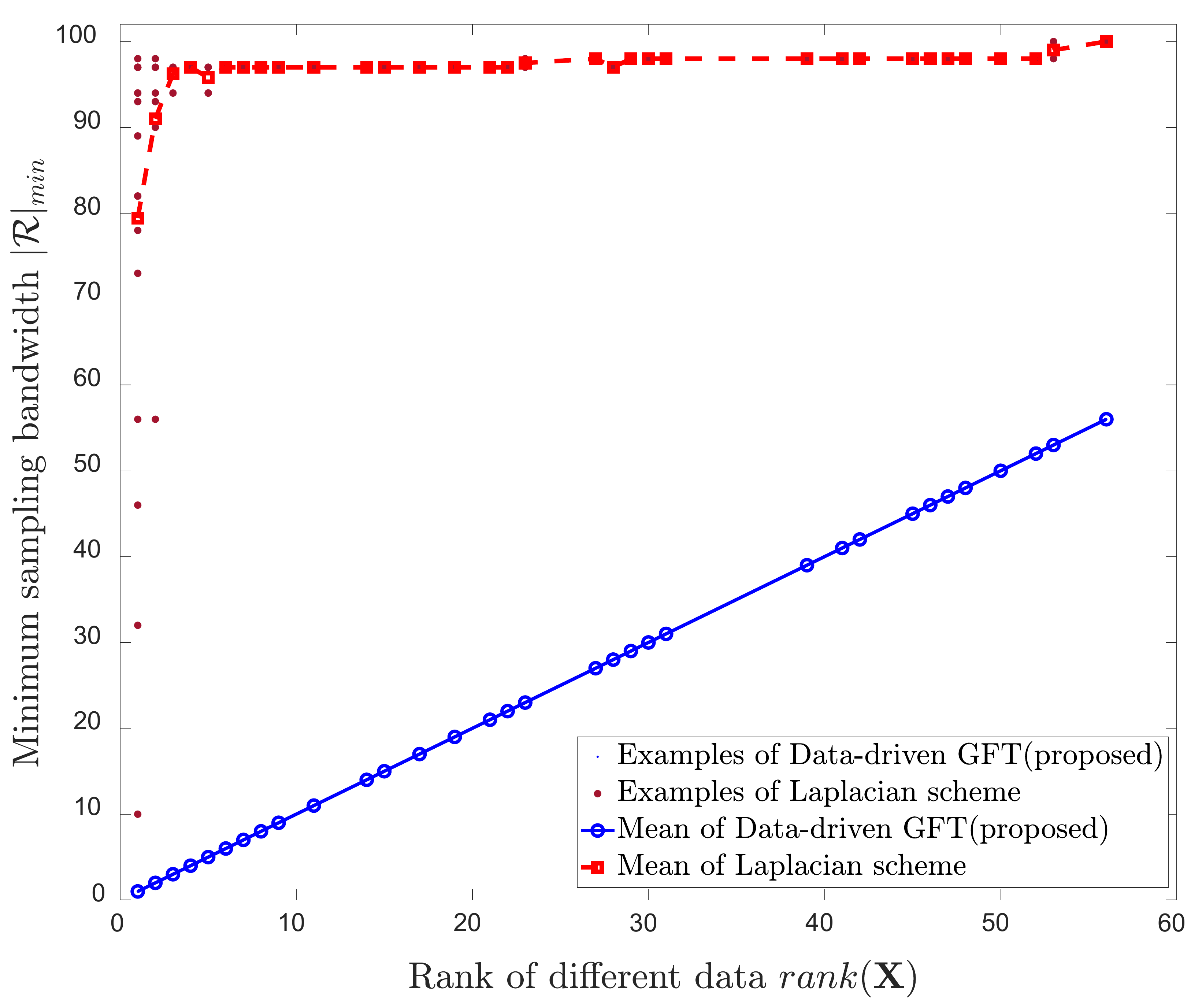}}
\hfil
\subfloat[]{\includegraphics[width=3.5in]{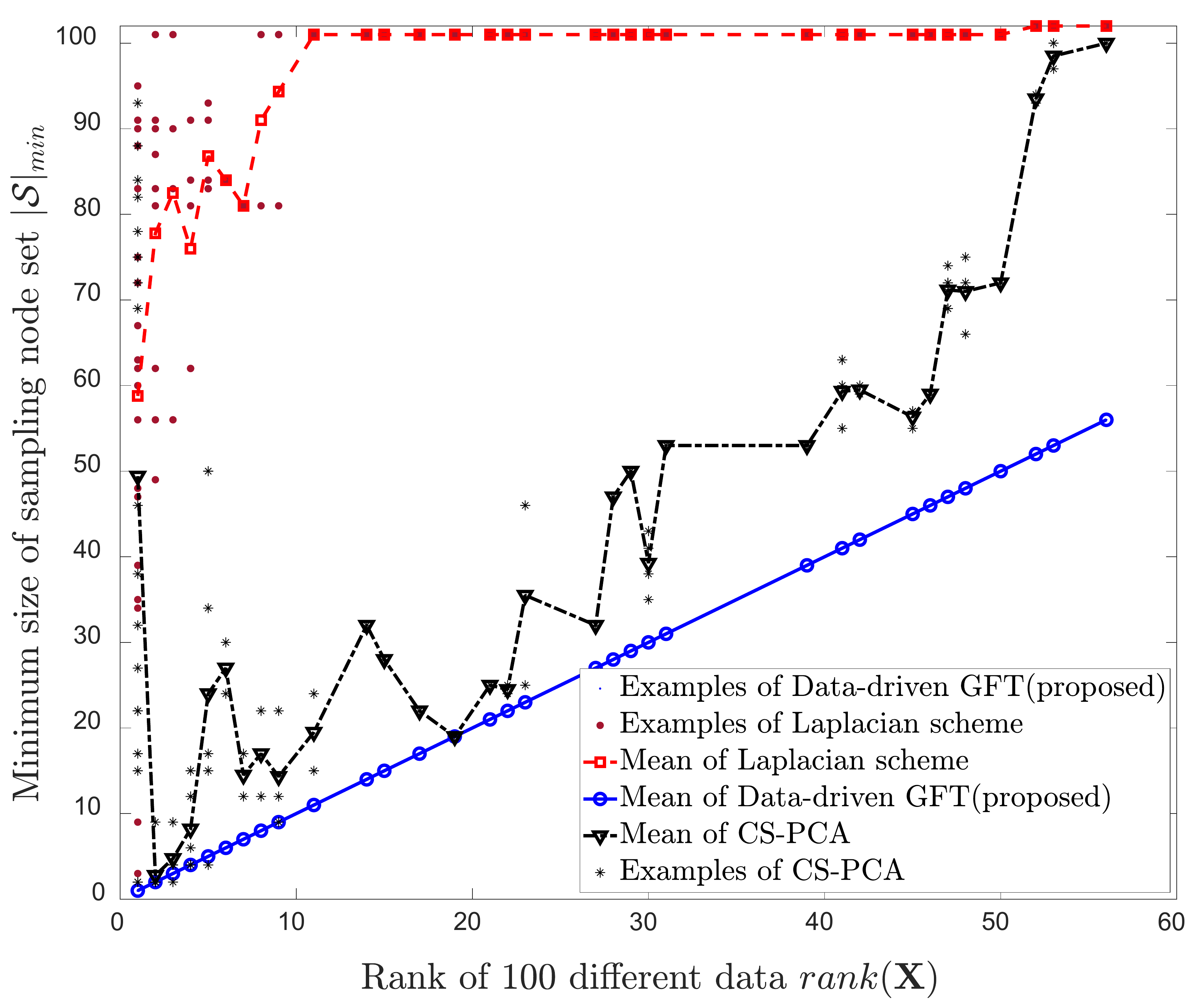}}
\caption{Minimum size of the sampling node set $|\mathcal{S}|_{\text{min}}$ such that RMSE$<10^{-8}$ for different 100 data, where x-coordinate gives the ranks of 100 data as $rank(\mathbf{X})$, while y-coordinate illustrates the $|\mathcal{S}|_{\text{min}}$. We compute the mean of different data with a same rank. It is seen that $|\mathcal{S}|_{\text{min}}$ from the proposed method always keeps at its minimum value as $|\mathcal{S}|_{\text{min}}=rank(\mathbf{X})$, greatly smaller than the scheme based on the traditional Laplacian operator, and the compressed sensing scheme. }
\label{set}
\end{figure*}

The performance comparison between our proposed sampling method, the sampling based on Laplacian operator, and the compressed sensing scheme is illustrated in Fig. \ref{comparison}-\ref{set}.

In Fig. \ref{comparison}(a), x-coordinate represents the frequency index from different domains\footnote{We here list the frequency indices from the proposed data-driven GFT operator domain, the Laplacian operator domain, and the PCA operator domain from the compressed sensing in the same x-coordinate, as they all have $N=102$ discrete frequencies. }. y-coordinate gives the summation of magnitudes of the frequency response in each time-step, i.e., $\sum_{k=1}^K|\mathbf{\mathbf{x}}_k|$. We can observe that the frequency response concentrates on the low-frequency area (i.e., $\mathcal{R}=\{1,\cdots,r\}$ with $r=41$) when using the proposed GFT operator, as opposed to the them using PCA operator and Laplacian operator respectively. This is because the Laplacian operator considers only the topology properties, and therefore cannot ensure the the networked data with time-varying dynamics $\mathbf{X}$ being bandlimited. Also, the perofrmance of the PCA operator is limited, given its overlook of the topology information. In contrast, our proposed data-driven GFT operator combines both the data and the topology properties, thereby capable of making $\tilde{\mathbf{X}}$ inside the low-frequency area $\mathcal{R}=\{1,\cdots,r\}$. As we mentioned before, this low-frequency characteristic with respect to the proposed GFT operator enables the selection of sampling nodes $\mathcal{S}$, which is shown in Fig. \ref{comparison}(b).

Fig. \ref{comparison}(b) presents recovery performance of three schemes with the changes of the size of the sampling node set $|\mathcal{S}|$. It is easily seen that as $|\mathcal{S}|$ increases, the RMSEs from all schemes decrease, due to the fact that a larger $|\mathcal{S}|$ can embrace larger amounts of samples, thereby leading to a better data recovery. Secondly, it is noteworthy that the RMSE of the proposed method decreases till $|\mathcal{S}|$ reaches the rank of the data, i.e., $|\mathcal{S}|=rank(\mathbf{X})=40$, and then converges to a constant (e.g., nearly $10^{-8}$ close to $0$) as $|\mathcal{S}|>rank(\mathbf{X})=40$. By contrast, the RMSE from other two methods decreases slowly, and can reach a perfect recovery (i.e., RMSE$\approx 0$) only when $|\mathcal{S}|$ approaches to $N=102$. This suggests that by relying on the proposed sampling method, we can use at least $|\mathcal{S}|=rank(\mathbf{X})$ nodes to sample and fully recover the networked dynamic data $\mathbf{X}$, which is greatly smaller than the value of the sampling scheme based on the traditional Laplacian operator, and the one based on the compressed sensing. The reason is that the proposed GFT operator is capable of transforming the data $\mathbf{X}$ into an upper triangular matrix with rank $r=rank(\mathbf{X})$, therefore we can use the first $r$-row of its GFT signal $\tilde{\mathbf{X}}$ to characterize $\mathbf{X}$. In this view, by selecting $|\mathcal{S}|\geq r$ rows from the GFT operator such that Eq. (\ref{selectnew}), we can ensure the fully recovery via Eq. (\ref{recover}).

Then, in order to demonstrate the robustness of our method, we measure the minimum sampling bandwidth, denoted as $\mathcal{R}_{\text{min}}$, and the minimum size of the sampling node set, denoted as $|\mathcal{S}|_{\text{min}}$ such that RMSE$<10^{-8}$ via 100 different data. In Fig. \ref{set}(a)-(b), the x-coordinate represents the ranks of different data, while y-coordinate present $\mathcal{R}_{\text{min}}$ and $|\mathcal{S}|_{\text{min}}$ respectively. We can firstly observe that with the increase of the rank of data, $\mathcal{R}_{\text{min}}$ and $|\mathcal{S}|_{\text{min}}$ of all schemes grow, which validates our theory that $|\mathcal{S}|\geq rank(\mathbf{X})$. More intriguingly, we can see that $|\mathcal{R}|_{\text{min}}$ and $|\mathcal{S}|_{\text{min}}$ from the proposed method always take their minimum value (i.e., $|\mathcal{R}|_{\text{min}}=rank(\mathbf{X})$ $|\mathcal{S}|_{\text{min}}=rank(\mathbf{X})$), which are greatly lower than the vlues used by the Laplacian scheme and the compressed sensing method. This suggests the robustness of our method in dealing with different dynamic data. The advantage of our scheme is alo attributed to the data-driven GFT operator, with respect to which the data $\mathbf{X}$ is $r$-bandlimited on only the frequencies indexed by $\{1,\cdots,rank(\mathbf{X})\}$, and therefore the fully recovery can be reached with $\mathcal{R}$ and $\mathcal{S}$ such that Eq. (\ref{15}) and Eq. (\ref{selectnew}) is satisfied. \\

\section{Conclusions and Discussion}

Water Distribution Networks (WDNs) are critical infrastructures that ensure safe drinking water. One of the major threats is the accidental or intentional injection of pollution in the system. Such threats, if not promptly detected, rapidly spreads in the whole system, affecting end-users. To contain the contamination and protect the population, it is fundamental to measure and predict the spread of the pollution in WDNs.

An open challenge is how to collect the minimum volume of data at critical junctions in order to infer the spread process across the rest of the network. Whilst numerical approaches through multi-objective optimisation and sensitivity analysis are well studied, they do not yield theoretical insights and are difficult to scale to larger networks and complex dynamics. On the other hand, graph theoretic approaches only consider the topology (e.g. Laplacian spectra) and do not factor in the essential dynamics.

In this work, we introduce a novel Graph Fourier Transform (GFT) to optimally sample junctions (nodes) in dynamic WDNs. The proposed GFT allows us to fully recover the full network dynamics using a subset of data sampled at critical nodes. This technique exploits the low rank property of the WDN dynamics, and offers attractive performance improvements over existing numerical optimisation, compressed sensing (CS), and graph theoretic approaches. Our results show that, on average, with nearly 30-40\% of the junctions monitored, we are able to fully recover the dynamics of the whole network. The framework is useful beyond the application of WDNs and can be applied to a variety of infrastructure sensing for digital twin modeling. \\

\textbf{Contributions: }
Z.K. developed the optimal sensing framework and conducted the analysis. W.G. and Z.K. developed the idea of the paper. A.P. simulated the water pollution dynamics. W.C. and J.M. provided guidance on compressed sensing comparative work. G.F. provided guidance on WDN modeling and comparative work on numerical optimisation. I.G. provided guidance on the problem context and impact pathway. Z.K., A.P., and W.G. wrote the paper. \\

\textbf{Acknowledgements:} The authors (A.P. \& W.G.) acknowledge funding from the Lloyd's Register Foundation's Programme for Data-Centric Engineering at The Alan Turing Institute. The authors (A.P., G.F. \& W.G.) acknowledge funding from The Alan Turing Institute under the EPSRC grant EP/N510129/1. The author (G.F.) acknowledge funding from EPSRC BRIM: Building Resilience Into risk Management (EP/N010329/1). \\
The authors acknowledge Microsoft Corporation for providing cloud resources on Microsoft Azure. \\

\bibliographystyle{IEEEtran}
\bibliography{IEEEabrv,bibfile}

\end{document}